\documentclass[aps,prb,twocolumn,superscriptaddress]{revtex4-2}
\usepackage{graphicx}
\usepackage{epstopdf}
\usepackage{amsmath}
\usepackage{appendix}
\usepackage{color}
\usepackage[colorlinks,linkcolor=blue,anchorcolor=blue,citecolor=blue,urlcolor=blue]{hyperref}
\usepackage{orcidlink}
\usepackage{mathtools} 
\usepackage{amssymb}
\usepackage{amsthm}

\begin{document}

\title{Intertwined charge density wave, tunable anti-dome superconductivity, and topological states in kagome metal VSn}

\author{Shu-Xiang Qiao\orcidlink{0009-0008-7092-3333}} 
\affiliation{School of Physics and Physical Engineering, Qufu Normal University, Qufu 273165, China}

\author{Ya-Ping Li}
\affiliation{School of Physics and Physical Engineering, Qufu Normal University, Qufu 273165, China}

\author{Jie Zhang}
\affiliation{School of Physics and Physical Engineering, Qufu Normal University, Qufu 273165, China}

\author{Yi Wan}
\affiliation{School of Physics and Physical Engineering, Qufu Normal University, Qufu 273165, China}

\author{Na Jiao}
\affiliation{School of Physics and Physical Engineering, Qufu Normal University, Qufu 273165, China}

\author{Meng-Meng Zheng}
\affiliation{School of Physics and Physical Engineering, Qufu Normal University, Qufu 273165, China}

\author{Hong-Yan Lu\orcidlink{0000-0003-4715-7489}}\email[E-mail: ]{hylu@qfnu.edu.cn} 
\affiliation{School of Physics and Physical Engineering, Qufu Normal University, Qufu 273165, China}

\author{Ping Zhang}\email[E-mail: ]{zhang$\_$ping@iapcm.ac.cn} 
\affiliation{School of Physics and Physical Engineering, Qufu Normal University, Qufu 273165, China}
\affiliation{Institute of Applied Physics and Computational Mathematics, Beijing 100088, China}

\begin{abstract}  	
 These years, kagome materials with 1:1 stoichiometry have garnered increasing attention, among which FeSn, CoSn, and FeGe have been the focus of current studies. However, all of them are antiferromagnetic, thereby hindering the observation of superconductivity and other novel physical properties. Here, we predict a novel 1:1 kagome metal VSn, which is an intrinsic charge density wave (CDW) material. Interestingly, with increasing pressure or doping concentration, the CDW order is progressively suppressed, followed by the emergence of superconductivity characterized by a non-monotonic transition temperature that exhibits a rare anti-dome-shaped dependence. Above a critical threshold, a reentrance of the CDW phase occurs. The anti-dome superconductivity originates from the first hardening and then softening of phonon modes, together with band reconstruction. Crucially, VSn retains nontrivial topological properties across the entire superconducting regime, a feature of paramount importance for realizing robust topological superconductivity. These intertwined CDW, superconductivity, and topological phenomena elucidate the correlations among multiple quantum states in VSn. Therefore, this research paves the way for for designing 1:1 kagome superconducting topological metals and establishes a platform for exploring the interplay of multiple phases in kagome systems.
\end{abstract} 

\maketitle

The kagome lattice, consisting of corner-sharing triangles, has emerged as an ideal platform for exploring novel quantum phenomena in condensed matter physics owing to its intrinsic geometric frustration \cite{Yinjx,zongshu1,zongshu2,zongshu3,zongshu4}. This unique lattice structure gives rise to the coexistence of key features in the electronic band structure, including flat bands \cite{Flatband1}, Dirac points \cite{Dirac}, and Van Hove singularities (VHSs) \cite{VHS1,VHS2,VHS3}. These characteristics enable a variety of emergent physical phenomena, including spin-liquid states \cite{spin-liquid1,spin-liquid2}, magnetic topological phases \cite{magnetic-topological1,fesn}, chiral phenomena \cite{chiral1,chiral2,chiral3}, anomalous Hall effects \cite{anhall1,anhall2}, charge density wave (CDW) order \cite{sijianguo}, and superconductivity \cite{sc1,sc2,xiaopengcheng,yangl,mpd5}.

In recent years, extensive theoretical and experimental efforts have led to the prediction and successful synthesis of numerous kagome-lattice materials, including the 135-type \cite{sijianguo,135-exp} and 166-type kagome compounds \cite{ka166,166-exp1,166-exp2}, among others \cite{fesn,anhall2,fe3sn21,fe3sn22}. These materials exhibit substantial diversity in atomic composition and geometric arrangement, leading to markedly distinct electronic properties. Notably, many kagome materials have been confirmed to exhibit CDW order and display diverse CDW-superconductivity phase diagrams under external pressure or doping. CsV$_{3}$Sb$_{5}$ has been experimentally \cite{135-exp,csvsb1} and theoretically \cite{sijianguo} verified to possess CDW properties and exhibit superconductivity under pressure. In addition, shallow $M$-shaped superconducting domains have also been observed in RbV$_{3}$Sb$_{5}$, with weak competition between CDW and superconductivity \cite{rbvsb}. The topology and weak superconducting behavior of 166 type kagome metal Y$T$$_{6}$Sn$_{6}$ ($T$ = V, Nb, Ta) have been theoretically predicted \cite{ka166}. Recently, kagome materials with a 1:1 stoichiometric ratio have garnered increasing attention. They include kagome lattice, hexagonal lattice, and triangular lattice. Among them, current experimental and theoretical studies have primarily focused on FeSn and CoSn \cite{fe3sn22,fesn,fesn2,cosn}, but both of them are antiferromagnetic, thereby hindering the observation of CDW or superconductivity. Therefore, the search for other materials in this kagome family and the study of their rich physical properties is necessary and will broaden the family of kagome materials exhibiting correlated quantum phenomena.

High pressure and doping are two commonly used tuning measures in condensed matter systems. As a clean tuning method, high pressure can continuously modify the crystal structure and electronic band structure without introducing chemical impurities. Hole doping can tune the Fermi level within the electronic band structure by introducing additional charge carriers, thereby inducing diverse quantum phase transitions. These two tuning approaches have been widely employed to elucidate the interplay between superconductivity and CDW order \cite{sijianguo,csvsb1,csvsb3,rbvsb}. Therefore, applying high pressure and doping techniques in kagome-lattice systems is crucial for exploring the phase diagrams and topological aspects associated with superconducting and CDW phases.

\begin{figure*}
	\centering
	\includegraphics[width=7.5cm]{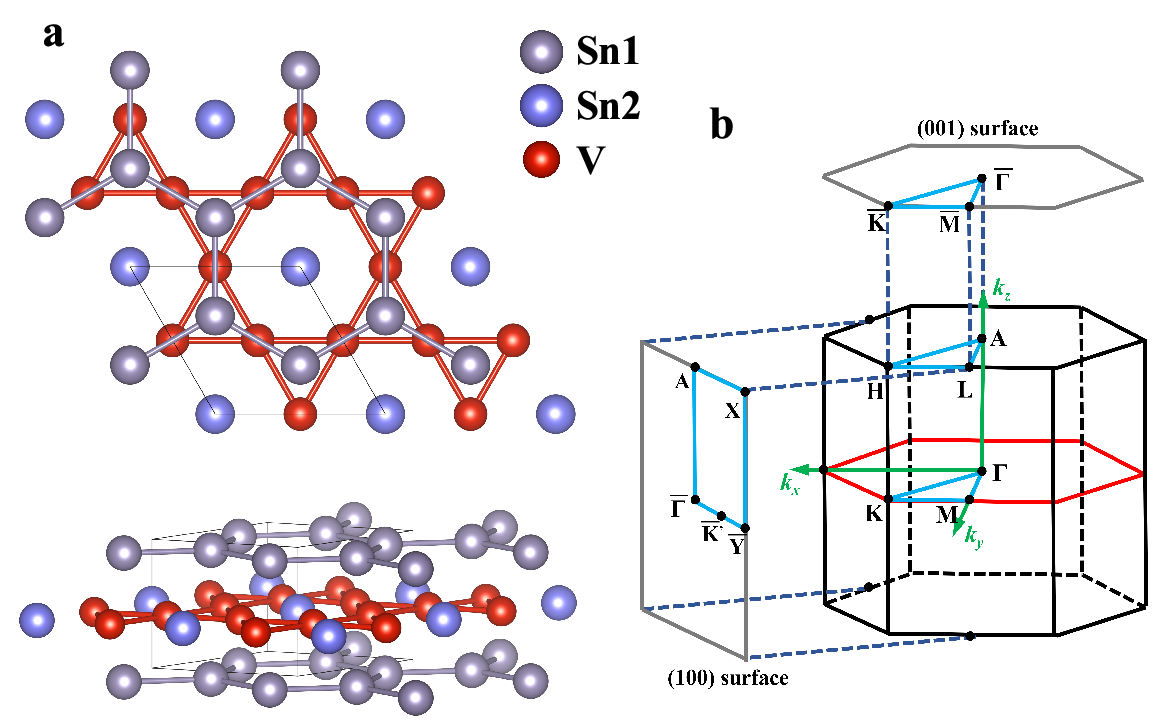} \includegraphics[width=8cm]{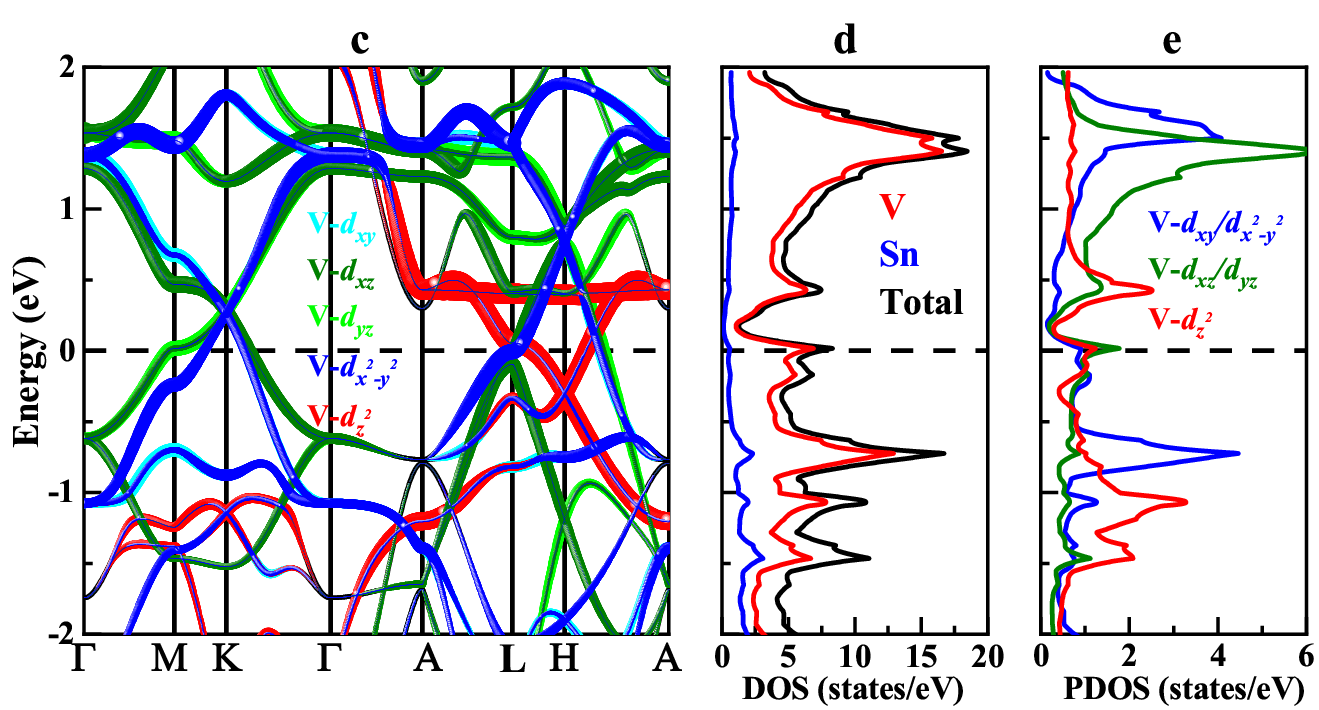}
	\includegraphics[width=9cm]{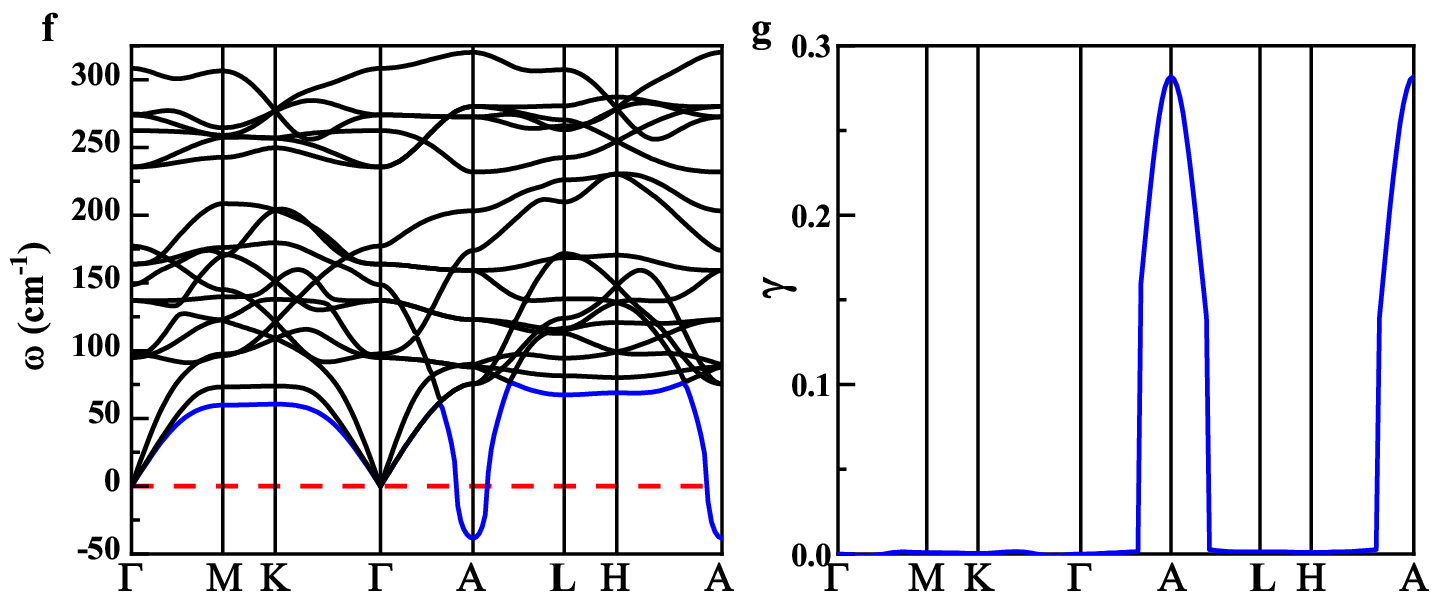}   \includegraphics[width=6.7cm]{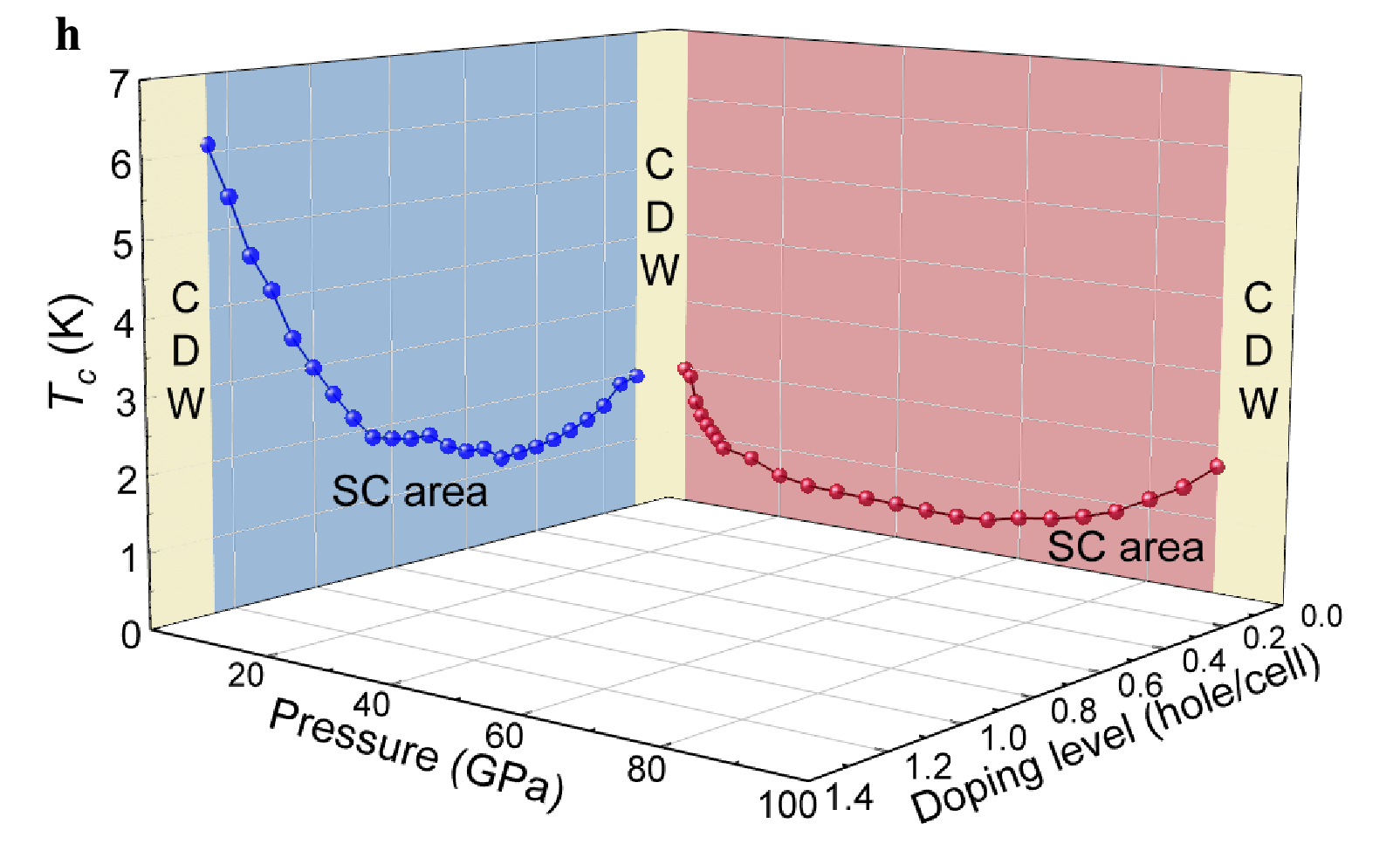}
	\caption{ \textbf{The crystal structure, Brillouin zone, electronic structures, CDW property, and phase diagram of VSn.} \textbf{a} Top and side views of VSn. \textbf{b} Bulk Brillouin zone and the high-symmetry path used in band structure calculations. The project (001) and (100) surfaces are also plotted. \textbf{c} Orbital-projected electronic band structure without SOC, \textbf{d} total DOS, \textbf{e} PDOS of the pristine VSn. \textbf{f} Phonon dispersions and \textbf{g} phonon linewidth $\gamma$ of the lowest phonon branch of VSn under ambient pressure. \textbf{h} The evolutions of $T_{c}$ of VSn as functions of pressure or hole doping.}
	\label{1}
\end{figure*}

In this work, utilizing first-principles calculations, we report the intrinsic CDW order of kagome metal VSn and reveal the anti-dome shaped superconducting phase diagram and the topological properties under pressure or hole doping. The CDW in VSn is driven by electron-phonon coupling (EPC) and can be suppressed by pressure or hole doping. Then, superconductivity appears with superconducting transition temperature ($T_{c}$) exhibiting an unusual anti-dome dependence under external tuning, arising from the combined effects of changing phonon modes and band reconstruction. Notably, within the superconducting range, VSn retains nontrivial topological characteristics, as confirmed by the calculated $\mathbb{Z}_{2}$ invariants and well-defined surface states. Thus, this study reveals the intertwined behavior of CDW, superconductivity, and topology in kagome metal VSn, providing valuable insights for exploring the interplay and phase evolution of these quantum states under different tuning techniques.

\textbf{Results and discussions}

\textbf{Crystal and electronic structures}

The crystal structure of VSn is shown in Fig. \ref{1}, which crystallizes in the hexagonal space group $P6/mmm$ (No. 191). The optimized lattice constants of VSn are $a$ = $b$ = 5.45 $\textmd{\AA}$ and $c$ = 4.64 $\textmd{\AA}$. The primitive cell contains three V atoms and three Sn atoms, corresponding to a 1:1 ratio. The Sn atoms occupy two distinct crystallographic sites, denoted as Sn1 and Sn2. As shown in Fig. \ref{1}a, the VSn structure consists of two distinct types of layers stacked along the $c$-axis: the Sn1 layer and the V-Sn2 layer. In the side view of the crystal structure (Fig. \ref{1}b), the V atoms form a kagome lattice in the $ab$ plane, while the Sn2 atoms constitute a triangular lattice within the same layer. The Sn1 atoms form a honeycomb network located between the kagome layers. This structural arrangement combines kagome, triangular, and hexagonal motifs, which may give rise to intriguing physical properties.

This structure adopts the FeSn structure as the parent phase, which has been experimentally synthesized and characterized using angle resolved photoelectron spectroscopy (ARPES) and de Haas van Alphen quantum oscillations \cite{fesn,fesn2}. However, FeSn is magnetic and does not exhibit superconductivity. Therefore, we intend to design new superconducting and CDW materials by substituting Fe with other transition-metal elements. Atomic substitution is a widely employed strategy for designing and synthesizing new materials. For example, CsAg$_{5}$Te$_{3}$ \cite{CsAg5Te3} has been reported as a thermoelectric material with ultralow thermal conductivity. Similarly, CsCu$_{5}$S$_{3}$ \cite{CsCu5S3} and CsCu$_{5}$Se$_{3}$ \cite{CsCu5Se3} with the same crystal structure are successfully synthesized via elemental substitution.

Figure \ref{1}c shows the band structure of VSn, which exhibits prominent kagome-lattice features near the Fermi level, including the Dirac points at the $K$ point and the flat band along the $A$-$L$-$H$-$A$ path. From the element projected band structure (Fig. \ref{1}c) and orbital projected density of states (PDOS) (Figs. \ref{1}d and \ref{1}e), the flat band states are found to originate primarily from the V-$d_{z^2}$ orbitals, whereas the electronic states near the Fermi level are dominated by the V-$d_{xz}$/$d_{yz}$ orbitals. As shown in Fig. \ref{1}c, four VHSs are located around the $M$ point in kagome VSn. From top to bottom, the first VHS mainly originates from the V-$d_{xy}$/$d_{x^2-y^2}$ orbitals, while the second arises from the V-$d_{xz}$ orbital. The third singularity, located at the Fermi level, is dominated by the V-$d_{xz}$/$d_{yz}$ orbitals and plays a crucial role in superconductivity \cite{vhs}. The fourth one is primarily associated with the V-$d_{x^2-y^2}$ orbital. In addition, multiple Dirac points are observed around the $K$ point, primarily derived from the V-$d_{xy}$/$d_{x^2-y^2}$ and $d_{xz}$/$d_{yz}$ orbitals, and they are located close to the Fermi level.

\textbf{CDW property}

As shown in Fig. \ref{1}f, a pronounced Kohn anomaly and an accompanying imaginary phonon frequency appear at the high-symmetry point $A$ in the phonon spectrum of VSn, which is a characteristic feature of CDW materials. Several mechanisms have been proposed to explain the formation of CDWs, among which the most widely studied and recognized are Fermi surface nesting (FSN) \cite{fsn1,fsn2}, momentum dependent EPC \cite{epc1,epc2,epc3}, and exciton condensation \cite{exciton1,exciton2}. Based on their formation mechanisms, CDWs can be classified into three types \cite{CDWtype1,CDWtype2}. Type-I CDW originates from quasi one-dimensional (1D) systems exhibiting FSN, while Type-II CDW arises primarily from strong EPC rather than nesting effects. Systems that do not fit into these two categories are classified as Type-III CDW. The FSN is generally responsible for the emergence of CDWs in quasi 1D and 2D materials. Exciton condensation, on the other hand, is more common in certain semimetallic and insulating systems \cite{exciton1,exciton2}. However, as shown in Fig. \ref{1}c, VSn possesses metallic characteristics, therefore, its CDW cannot be attributed to exciton condensation. Then, we examine the origin of the CDW in kagome VSn with the effect of momentum dependent EPC. As described in Fig. \ref{1}g, the phonon linewidth $\gamma$ of the lowest phonon branch of VSn is calculated, which is directly related to the strength of EPC. A pronounced peak is observed at the $A$ point, corresponding precisely to the position of the imaginary frequency in the phonon spectrum. Therefore, we conclude that EPC is the primary reason behind the formation of CDW in VSn.

\textbf{Controlled superconducting phase diagram}

The relationship between CDW and superconductivity has attracted extensive research attention in condensed matter physics. In kagome materials such as CsV$_{3}$Sb$_{5}$ and RbV$_{3}$Sb$_{5}$, the interplay between CDW order and superconductivity under pressure or doping has been extensively investigated experimentally \cite{csvsb1,csvsb2,csvsb3,rbvsb}. In this work, we focus on the evolution of CDW order and the superconducting $T_c$ of VSn under pressure or hole doping.

Figure \ref{1}h presents the phase diagrams of the CDW and superconducting $T_{c}$ of VSn under pressure or hole doping. The results indicate that with increasing pressure or hole doping, the CDW order is initially suppressed, followed by the emergence of superconductivity, and eventually a reappearance of the CDW phase at higher regulation levels. Once the CDW is suppressed, the superconducting $T_c$ displays an uncommon anti-dome behavior, first decreasing and then increasing, which differs from the typical dome shaped or double dome superconducting phase diagrams observed in CsV$_{3}$Sb$_{5}$ \cite{csvsb1,csvsb2} and RbV$_{3}$Sb$_{5}$ \cite{rbvsb}. Under external pressure, the CDW phase is suppressed at 3 GPa. With further compression, $T_c$ decreases from 2 K at 3 GPa to a minimum of 0.45 K at 50 GPa, then gradually rises to 1.73 K at 90 GPa, as shown in Fig. \ref{1}h. Beyond 90 GPa, the system reenters the CDW phase. A similar trend is observed under hole doping. At 0.1 hole doping, the CDW is suppressed and superconductivity emerges with $T_{c}$ of 1.78 K. As the doping level increases, $T_{c}$ decreases to 1.17 K at 0.5 hole doping, then rises to a maximum of 6.11 K at 1.25 hole doping. Above 1.25 hole doping, the CDW phase again becomes energetically favorable. Therefore, VSn exhibits a tunable anti-dome superconducting phase diagram under pressure or hole doping, as summarized in Fig. \ref{1}h.

 \begin{table}
 	\caption{ Superconducting $T_{c}$, total EPC constant $\lambda$, the logarithmically averaged phonon frequency $\omega_{log}$, and DOS at the Fermi level $N(E_{F})$ of VSn under pressure or doping.}
 	\centering
 	\setlength{\tabcolsep}{6pt}
 	\renewcommand{\arraystretch}{1}
 	\begin{tabular}{ccccc}
 		\hline
 		\hline
 		Materials      & $T_{c}$ (K)    & $\lambda$  & $\omega_{log}$ (K) & $N(E_{F})$ \\ \hline
 		VSn (3 GPa)        & 2.00	        & 0.36	     & 236.91	          & 34.69	   \\ 
 		VSn (50 GPa)       & 0.45	        & 0.27	     & 280.40	          & 25.52	   \\ 
 		VSn (90 GPa)       & 1.73	        & 0.34	     & 238.59	          & 29.71	   \\ 
 		VSn (0.1 hole/cell)    & 1.78	        & 0.36	     & 238.31	          & 38.89	   \\ 
 		VSn (0.5 hole/cell)    & 1.17	        & 0.33	     & 251.96	          & 33.41	   \\ 
 		VSn (1.25 hole/cell)   & 6.11	        & 0.71	     & 117.88	          & 41.37	   \\ 
 		\hline \hline
 	\end{tabular}
 	\label{sc}
 \end{table}

\begin{figure*}
	\centering
	\includegraphics[width=16cm]{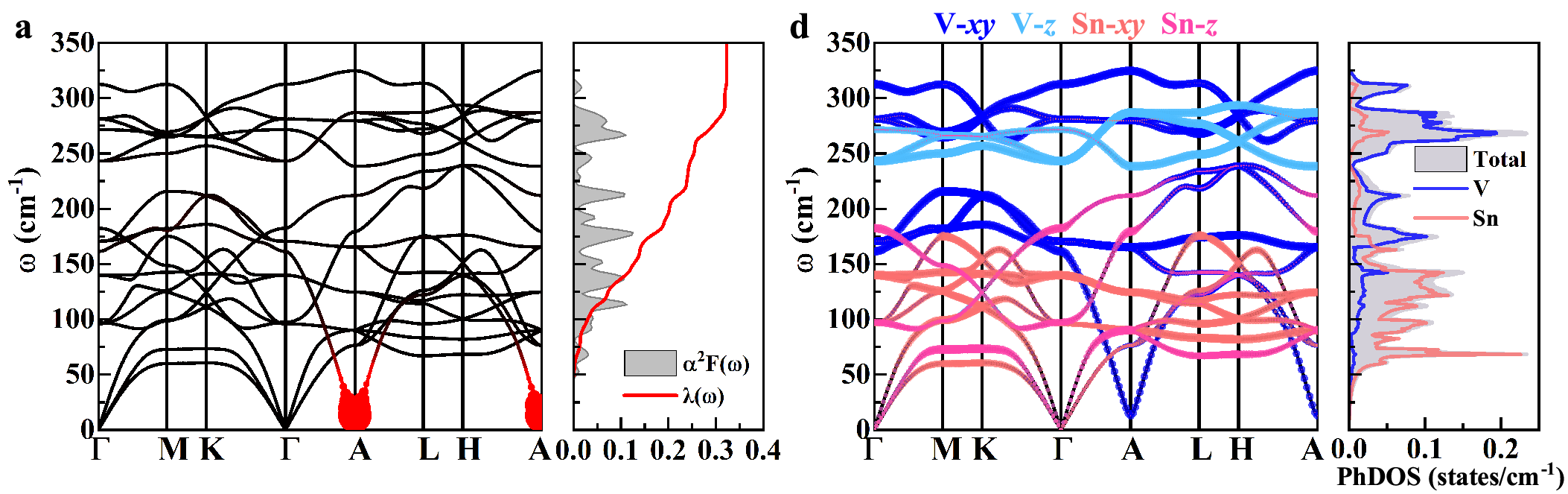}
	\includegraphics[width=16cm]{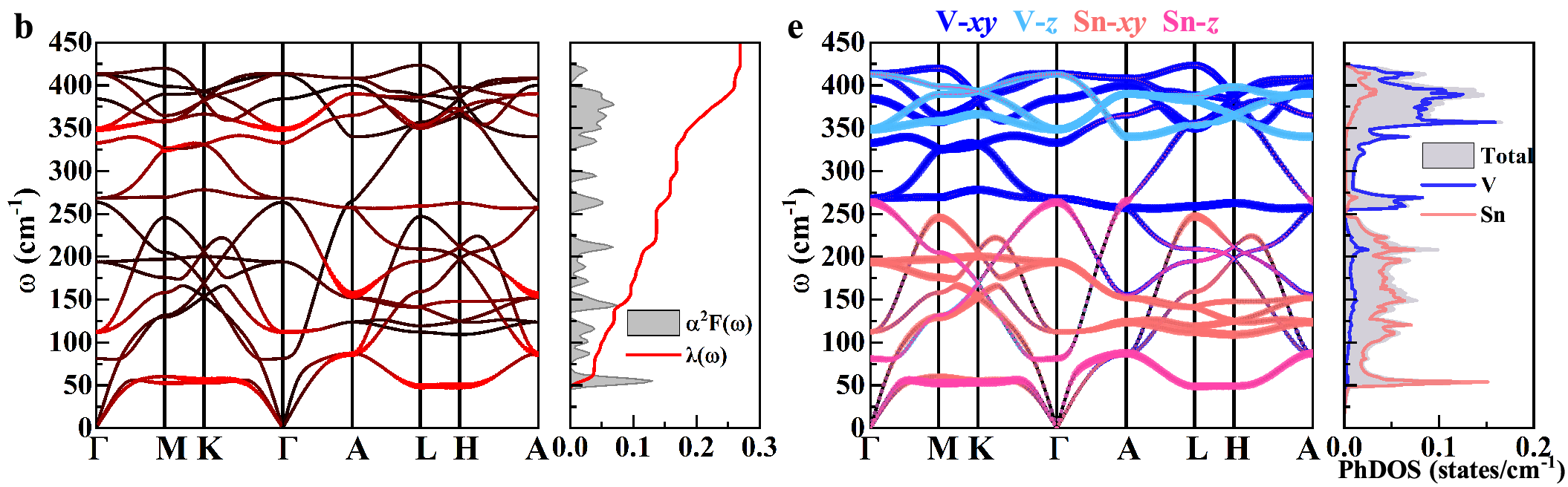}
	\includegraphics[width=16cm]{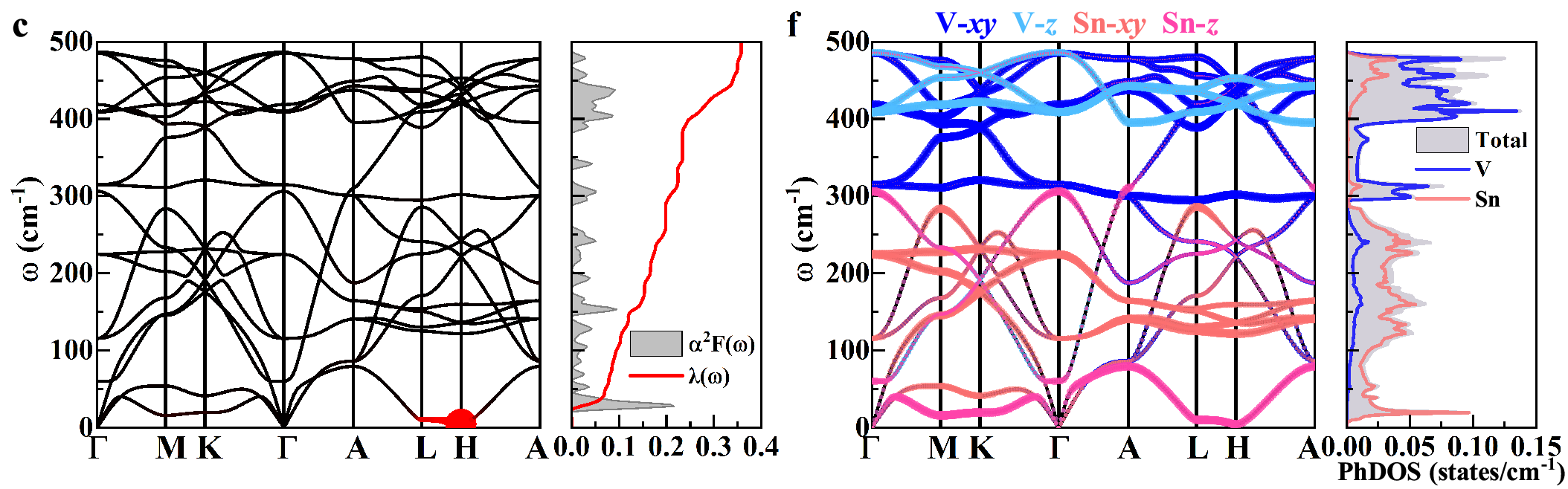}
	\caption{ \textbf{Superconductivity of VSn under pressure.} Phonon dispersion weighted by the magnitude of $\lambda_{\textbf{q}\nu}$ (EPC for phonon mode ${\textbf{q}\nu}$), Eliashberg spectral function $\alpha^{2}F (\omega)$, and EPC $\lambda (\omega)$ for VSn under pressures of \textbf{a} 3, \textbf{b} 50, and \textbf{c} 90 GPa. Phonon dispersion weighted by the vibration modes of each atom and PhDOS for VSn under pressures of \textbf{d} 3, \textbf{e} 50, and \textbf{f} 90 GPa.}
	\label{5}
\end{figure*}

\begin{figure*}
	\centering
	\includegraphics[width=16cm]{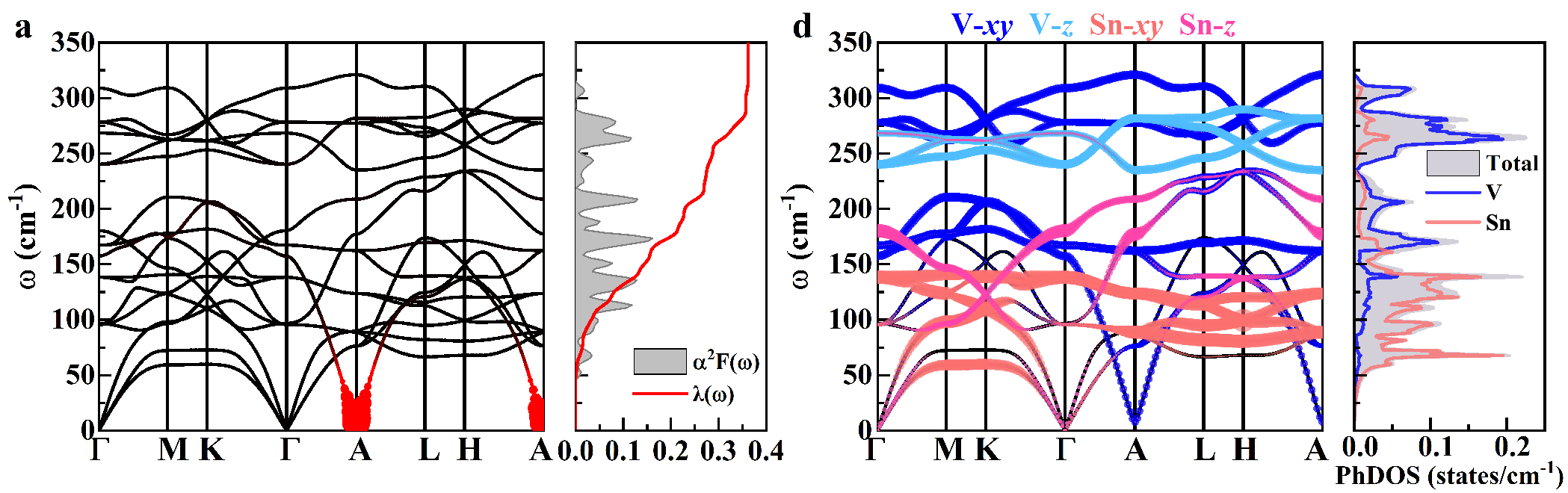}
	\includegraphics[width=16cm]{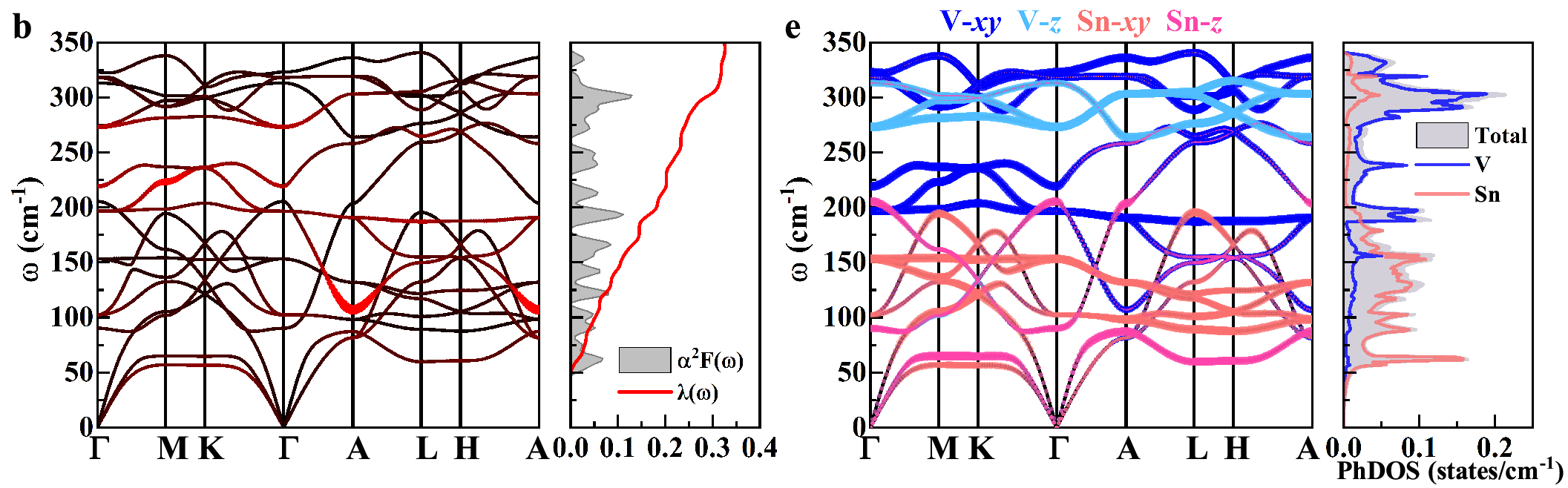}
	\includegraphics[width=16cm]{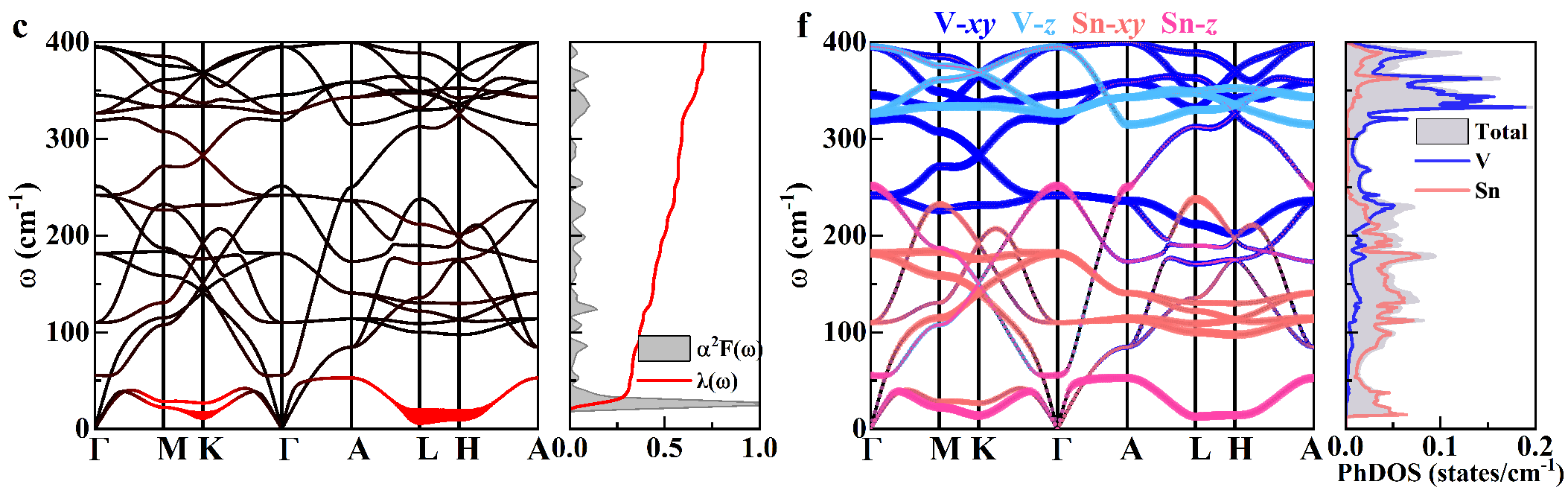}
	\caption{ \textbf{Superconductivity of doped VSn.} Phonon dispersion weighted by the magnitude of $\lambda_{\textbf{q}\nu}$ (EPC for phonon mode ${\textbf{q}\nu}$), Eliashberg spectral function $\alpha^{2}F (\omega)$, and EPC $\lambda (\omega)$ for VSn under doping levels of \textbf{a} 0.1, \textbf{b} 0.5, and \textbf{c} 1.25 holes/cell. Phonon dispersion weighted by the vibration modes of each atom and PhDOS for VSn under doping levels of \textbf{d} 0.1, \textbf{e} 0.5, and \textbf{f} 1.25 holes/cell.}
	\label{6}
\end{figure*}

According to Bardeen-Cooper-Schrieffer theory, superconductivity is intimately related to the strength of the EPC. Table \ref{sc} shows that the variations of the superconducting $T_c$ and the EPC constant $\lambda$ are generally consistent, both exhibiting a trend of first decreasing and then increasing. To further elucidate the superconducting phase diagram of VSn, we analyze the EPC calculation results at three representative pressures (3, 50, and 90 GPa) and three hole-doping concentrations (0.1, 0.5, and 1.25), as shown in Figs. \ref{5} and \ref{6}, respectively.

Figures \ref{5}a-\ref{5}c show the phonon dispersion weighted by the magnitude of $\lambda_{\textbf{q}\nu}$, Eliashberg spectral function $\alpha^{2}F (\omega)$ and EPC $\lambda (\omega)$ of VSn at pressures of 3, 50, and 90 GPa, respectively. At 3 GPa, the imaginary frequencies in the phonon spectrum disappear, indicating that the CDW is fully suppressed. As shown in Fig. \ref{5}a, a noticeable phonon mode softening occurs near the $A$ point, which enhances EPC. When the pressure is increased to 50 GPa (Fig. \ref{5}b), the soft mode at $A$ gradually hardens, leading to a reduction in $T_{c}$. As the pressure increases further to 90 GPa (Fig. \ref{5}c), the lowest phonon branches soften at the $L$ and $H$ points, resulting in enhanced EPC. When the pressure exceeds 90 GPa, imaginary frequencies reappear at the $M$, $L$, and $H$ points, as shown in Fig. S1, indicating that VSn becomes dynamically unstable and the CDW order re-emerges. Throughout this process, the EPC constant $\lambda$ first decreases and then increases, following the hardening of existing phonon modes and the softening of newly emerging ones, mirroring the variation in $T_c$.

Figures \ref{5}d-\ref{5}f show the phonon dispersion weighted by the vibration modes of each atom and phonon density of states (PhDOS) at pressures of 3, 50, and 90 GPa. To clarify the contributions of different vibrational modes to superconductivity, we compare the phonon dispersion weighted by EPC intensity in Figs. \ref{5}a-\ref{5}c with that weighted by atomic vibration modes in Figs. \ref{5}d-\ref{5}f. This comparison reveals that at 3 GPa, the EPC is predominantly contributed by the V-$xy$ plane vibrations. At 50 GPa, both V and Sn vibrations play comparable roles in EPC. As the pressure further increases to 90 GPa, the low-frequency Sn-$z$ plane vibrations make an increasingly significant contribution to superconductivity, with the softened phonon modes accounting for approximately 46.5 $\%$ of the total EPC. Therefore, from the perspective of EPC, the origin of the anti-dome superconducting phase diagram of VSn under pressure can be attributed to the evolution of phonon softening. With increasing pressure, the initially softened phonon modes gradually harden, while new soft modes emerge. During this process, the dominant EPC contribution shifts from V-$xy$ plane vibrations to Sn-$z$ plane vibrations.

\begin{figure*}
	\centering
	\includegraphics[width=16cm]{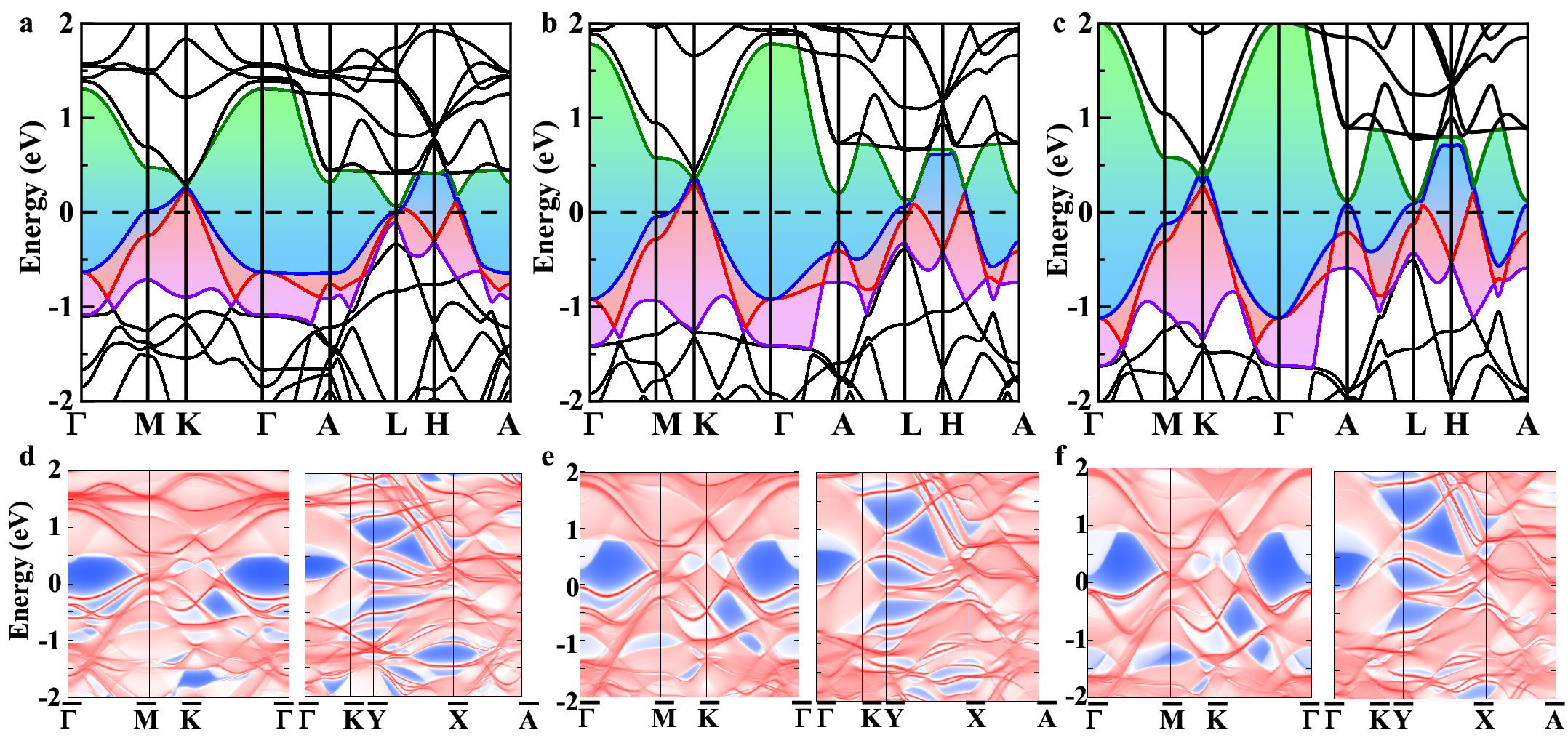}
	\includegraphics[width=16cm]{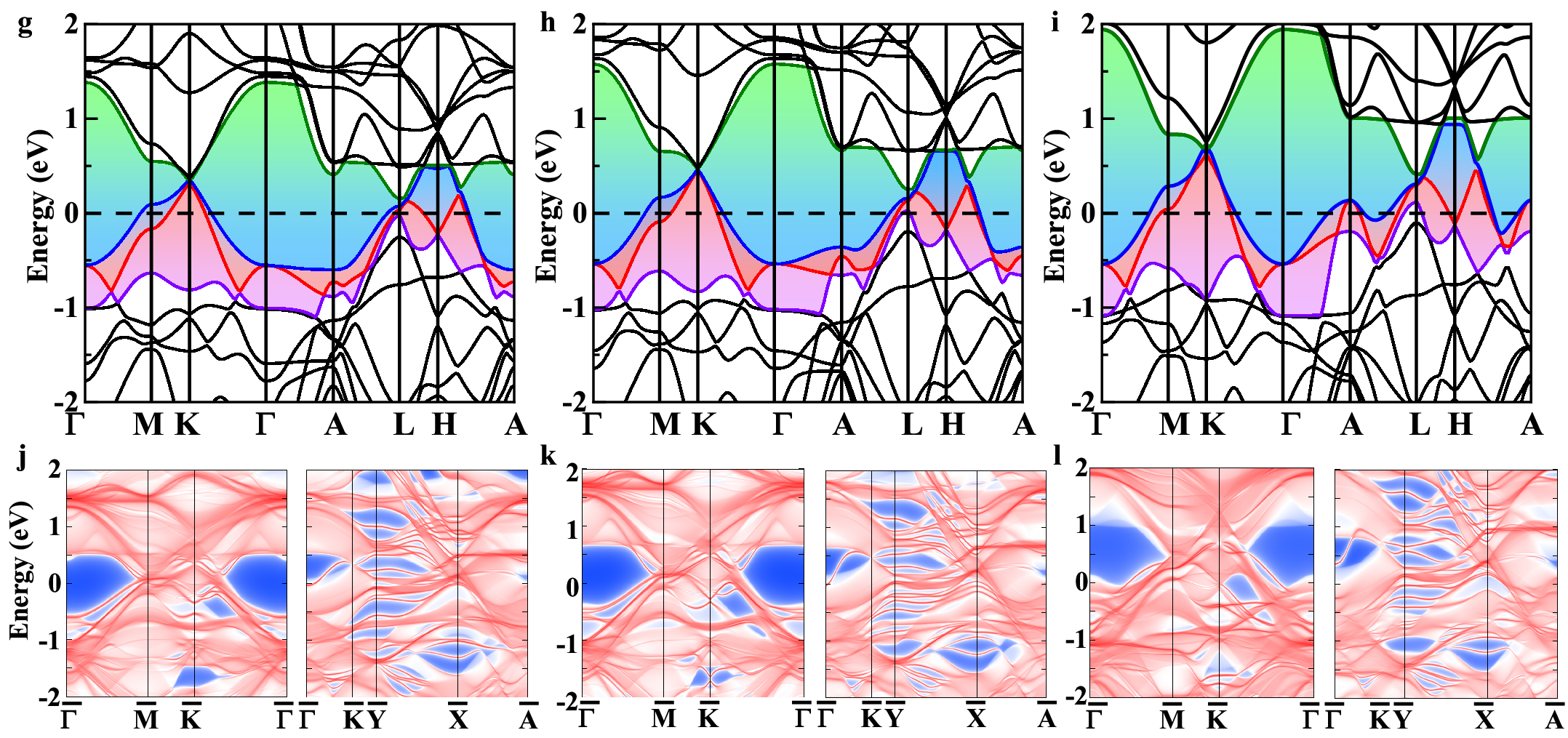}
	\caption{ \textbf{Topological properties of VSn under pressure and doping.} Band structures of VSn under pressures of \textbf{a} 3, \textbf{b} 50, \textbf{c} 90 GPa, and doping levels of \textbf{g} 0.1, \textbf{h} 0.5, \textbf{i} 1.25 holes/cell. Their corresponding (001) (left) and (100) (right) surface states are plotted in \textbf{d}, \textbf{e}, \textbf{f}, \textbf{j}, \textbf{k}, and \textbf{l}, respectively.}
	\label{7}
\end{figure*}

The calculated EPC results under three representative hole doping concentrations are shown in Fig. \ref{6}. Figures \ref{6}a-\ref{6}c display the phonon dispersion weighted by the magnitude of $\lambda_{\textbf{q}\nu}$, the Eliashberg spectral function $\alpha^{2}F(\omega)$, and the cumulative EPC function $\lambda(\omega)$ for 0.1, 0.5, and 1.25 hole doped VSn, respectively. Figures \ref{6}d-\ref{6}f present the phonon dispersion weighted by the vibration modes of each atom and PhDOS under the same doping levels. A comparative analysis reveals that the evolution of phonon modes under hole doping follows a trend similar to that observed under pressure. At 0.1 hole doping, the phonon spectrum of VSn becomes dynamically stable, indicating that the CDW order is suppressed. As the doping concentration increases to 1.25 holes, phonon instabilities reappear at the $M$, $L$, and $H$ points, signaling the reemergence of CDW order. Furthermore, the evolution of the softened phonon modes that are critical for superconductivity also exhibits behavior analogous to that under pressure. The initially softened phonons, dominated by V-$xy$ plane vibrations, gradually harden with increasing doping concentration, while new softened phonon modes primarily originated from Sn-$z$ plane vibrations emerge at the $L$ and $H$ points. This transition in the dominant vibrational contribution is the key factor underlying the anti-dome superconducting phase diagram observed in hole-doped VSn.

Therefore, VSn is an intrinsic CDW material whose CDW order can be suppressed by hydrostatic pressure or hole doping. As the pressure or doping concentration increase, VSn exhibits an anti-dome superconducting behavior. When the pressure or doping level exceeds a critical threshold, imaginary frequencies reappear, signaling the recovery of the CDW phase. Analysis of the EPC strength, which is directly related to superconductivity, reveals that the variation trend of the EPC constant $\lambda$ is consistent with that of $T_{c}$, both showing an anti-dome dependence. Further EPC calculations under three representative pressures and doping concentrations demonstrate that the formation of the anti-dome superconducting phase diagram originates from the hardening of softened phonon modes, followed by the emergence of new softened phonon modes. Therefore, this changing phonon softening plays a crucial role in shaping the anti-dome superconducting phase diagram of VSn.

Figures \ref{7}a-\ref{7}c present the band structures of VSn under 3, 50, and 90 GPa. By comparing the band structures at 3 and 50 GPa, it is found that the bandwidths of the two bands crossing the Fermi level (green and blue curves in Fig. \ref{7}b) become broader. The DOS at the Fermi level $N(E_{F})$ decreases from 34.69 at 3 GPa to 25.52 at 50 GPa, consistent with the decreasing trend of $T_{c}$ shown in Fig. \ref{1}h. This decrease primarily arises because the VHSs at the $M$ point shifts away from the Fermi level, reducing $N(E_{F})$. At 90 GPa, the blue energy band crosses the Fermi level near the $A$ point, clearly indicating a Lifshitz phase transition. This reconstruction results in an increased DOS at the Fermi level, as summarized in Table \ref{sc}. Therefore, we attribute the re-enhancement of superconductivity at high pressure to this pressure induced band reconstruction \cite{sijianguo}, which is accompanied by an increase in $N(E_{F})$. Similarly, the band structures calculated at different hole doping concentrations exhibit behavior analogous to that under pressure, as shown in Figs. \ref{7}g-\ref{7}i. At 0.5 hole doping, the VHS at the $M$ point shifts away from the Fermi level, reducing the DOS, consistent with the anti-dome superconducting behavior shown in Fig. \ref{1}h. When the doping concentration increases further to 1.25 holes/cell, a band reconstruction near the $A$ point enhances the DOS, thereby promoting superconductivity. Therefore, the band reconstruction near the $A$ point in hole doped VSn plays a crucial role in the evolution of its superconducting properties.

From the perspective of the electronic structure, it is found that the position of the VHSs and the band reconstruction have an impact on the DOS at the Fermi level. These changes in DOS play the important role in regulating $T_{c}$. In summary, combined the results of EPC and electronic structure analyses, we identify the formation mechanism of the anti-dome superconducting phase diagram. The evolution of the VHS, band reconstruction, and the breathing soft phonon modes are the key factors governing the superconductivity of VSn under varying pressure and doping.

\textbf{Topological properties under pressure or doping}

After elucidating the mechanism of the superconducting phase diagram in VSn, we further explore its topological states. The topological properties of VSn are investigated by analyzing the $\mathbb{Z}_{2}$ topological invariant and surface states \cite{topology}. As mentioned above, this material hosts Dirac points and flat band features, consistent with the typical characteristics of kagome materials that exhibit Dirac band crossings and VHSs \cite{chiral3,kagome3}. Figures \ref{7}a-\ref{7}c and \ref{7}g-\ref{7}i present the band structures of VSn under three representative pressures and hole doping concentrations. The VHS and flat bands are found to persist stably near the $M$ point and along the $A$-$L$-$H$-$A$ path, respectively. Upon including spin-orbit coupling (SOC), continuous energy gaps open throughout the Brillouin zone near the Fermi level. Similar to the kagome metal CsV$_{3}$Sb$_{5}$ \cite{sijianguo}, Y$T$$_{6}$Sn$_{5}$ \cite{ka166}, and $M$Pd$_{5}$ \cite{mpd5}, the VSn series materials preserve both spatial inversion and time-reversal symmetries.

Owing to the continuous energy gaps throughout the Brillouin zone, the $\mathbb{Z}_{2}$ topological index is calculated. The results reveal that VSn exhibits nontrivial topological properties under both pressure and hole doping. In VSn, the (001) surface exhibits easy cleavage characteristics due to weak interlayer interactions, whereas the (100) surface displays distinct electronic features. Therefore, we calculate the surface states for both the (001) and (100) surfaces, as shown in Figs. \ref{7}d-\ref{7}f and \ref{7}j-\ref{7}l. Analysis of the surface reveals numerous pronounced surface states near the Fermi level. In all six cases, these surface states are clearly located near the Fermi level and could be experimentally detected by ARPES measurements. Moreover, it can be observed that there are surface flat bands on the (100) surface, some of which hybridize with the bulk states. These complex surface states may give rise to intriguing transport phenomena and potential topological superconductivity.

Based on the above, VSn can be identified as a superconducting topological metal possessing both superconductivity and nontrivial topology, making it an ideal platform for realizing and studying topological superconductivity. Experimentally, superconductivity is often realized by doping topological insulators (such as Cu- or Nb-doped Bi$_{2}$Se$_{3}$ \cite{CuBiSe,NbBiSe} and Sn$_{1-x}$In$_{x}$Te \cite{SnTe}) or through the proximity effect in heterostructures \cite{tsc1,tsc2}. In contrast to these doped compounds and heterostructures, VSn exhibits intrinsic superconductivity and nontrivial topological properties, offering valuable insight for the design of new topological superconductors.

\textbf{Conclusion}

In this study, we have predicted kagome maetal VSn and systematically investigated the properities using first-principles calculations, unveiling its intrinsic CDW, anti-dome superconducting phase diagram, and topological properties under pressure or hole doping. This work reveals a three-phase interweaving of CDW, anti-dome superconductivity, and nontrivial topological order in kagome VSn. The intrinsic CDW originates from EPC and can be effectively suppressed by pressure or hole doping. A rare anti-dome superconducting behavior is identified, attributed to the combined effects of changing phonon modes and band reconstruction. Remarkably, within the superconducting range, VSn consistently maintains nontrivial topological properties, as confirmed by $\mathbb{Z}_{2}$ topological invariants and clearly surface states. This three-phase interweaving provides a crucial insight into the interplay among multiple correlated phenomena in VSn. Therefore, this research paves the way for for designing 1:1 kagome superconducting topological metals and establishes a platform for exploring the interplay of multiple phases in kagome systems.

\textbf{Methods}

All calculations in this paper are carried out in the framework of density functional theory (DFT), as implemented in the Vienna ab-initio simulation package (VASP) \cite{1} and the software QUANTUM-ESPRESSO (QE) \cite{2}. The exchange-correlation potentials are described with the generalized gradient approximation (GGA) with Perdew–Burke–Ernzerhof (PBE) parametrization \cite{3}. The projector augmented-wave (PAW) technique is chosen to measure the electron-ion interactions \cite{4}. The electronic structures are calculated by using VASP, the phonon and superconductivity properties are calculated by using QE. In the calculation, the wave function and charge density cutoff energy are 80 and 800 Ry, respectively. 6×6×8 $k$-point grid is used for charge self consistent calculation, and a 3×3×4 $q$-point grid for dynamical matrix. For DOS calculations, 18×18×24 $k$-point grid is used. The surface states are determined through the iterative Green's function implemented in the WANNIERTOOLS package \cite{5,6}, and its basis set depends on the maximum localized Wannier functions (MLWFs) \cite{7,8} from the VASP2WANNIER90 interfaces \cite{9}.

\textbf{Data availability}

The data used in this study are available from the corresponding authors upon request.

\textbf{Author contributions}

S. X. Q. performed all calculations, including structural screening, stability, electronic, topological, phononic, and superconducting properties. All authors including Y. P. L., J. Z., Y. W., N. J., M. M. Z., H. Y. L. and P. Z. contributed to the analysis and interpretation of the data. S. X. Q. wrote the manuscript with input from all coauthors. N. J. and H. Y. L. designed the project. H. Y. L. and P. Z. supervised the project. All authors had contributed to the revision of the manuscript.

\textbf{Competing interests}

The authors declare no competing interests.

\textbf{Acknowledgements}

 This work is supported by the National Natural Science Foundation of China (Grant No. 12074213), the National Key R$\&$D Program of China (Grant No. 2022YFA1403103), the Major Basic Program of Natural Science Foundation of Shandong Province (Grant No. ZR2021ZD01), and the Natural Science Foundation of Shandong Province (Grant No. ZR2023MA082).

\bibliography{references}

\end{document}